\documentclass[conference]{IEEEtran}
\usepackage{array}
\usepackage{cite}
\usepackage{amsmath,amssymb,amsfonts}
\usepackage{algorithmic}
\usepackage{graphicx}
\usepackage{textcomp}
\usepackage{xcolor}
\usepackage{url} 
\usepackage[top=0.77in, bottom=1.1in, left=0.65in, right=0.65in]{geometry}
\def\BibTeX{{\rm B\kern-.05em{\sc i\kern-.025em b}\kern-.08em
    T\kern-.1667em\lower.7ex\hbox{E}\kern-.125emX}}
\begin{document}

\title{Design and Analysis of Power Consumption Models for Open-RAN Architectures}

\author{\IEEEauthorblockN{Urooj Tariq, Rishu Raj, and Dan Kilper}\small{CONNECT Centre, Trinity College Dublin, Ireland; Email: \{tariqu, rajr, dan.kilper\}@tcd.ie}}
\maketitle

\begin{abstract} 
The open radio access network (O-RAN) Alliance developed an architecture and specifications for open and disaggregated cellular networks including many elements that are being widely adopted and implemented in both commercial and research networks. In this paper, we develop transaction-based power consumption models of a centralized O-RAN architecture based on commercial hardware and considering the full end-to-end data path from the radio unit to the data center. We focus on recent fanout limitations and early baseband processing requirements related to current implementations of O-RAN and assess the power consumption impact when baseband processing is employed at different centralization points in the network. Additionally, we explore how greater fanout and sharing deeper into the network impact the balance of processing and transmission. Low processing fanout restrictions motivate greater centralization of the processing. At the same time, allowing for more open radio units per open distributed unit will quickly increase the transmission capacity requirements and related energy use.

\end{abstract}


\section{Introduction}
\IEEEPARstart{C}{ommunication}\label{Section I}
 networks are evolving in response to emerging connectivity paradigms such as artificial intelligence (AI) native networks that provide greater functionality and reach. This evolution has been accompanied by important architectural changes allowing for greater centralization and virtualization of network functions. A greater reliance on virtualization has also enabled open networking standards such as the open radio access network (O-RAN) specification \cite{oran:online}. O-RAN allows for the use of network control and management applications that could be offered by third party vendors. It leverages open standards and interfaces to decouple hardware and software components. This enables network operators to mix and match solutions from different vendors, increasing flexibility and reducing dependency on a single supplier. The potential benefits include cost reduction, faster innovation through competition, easier scalability, and a more dynamic ecosystem that fosters collaboration.
 
Investigating the power consumption in O-RANs is essential for optimizing energy efficiency, reducing operational costs, and minimizing environmental impact, thereby making 5G/6G and future networks more sustainable and scalable. The O-RAN Alliance has identified energy-efficient network operation as a key early use case for O-RAN network management applications \cite{ORANVert0:online}. Additionally, efficient power management is crucial for meeting green network initiatives and reducing carbon footprints in modern telecommunications. As such many recent studies of energy use in O-RAN based networks have emerged. However, much of this work has focused on the radio access network portion of the network with little or no consideration for the overall end-to-end network energy use, from the radio units to the data center \cite{baliga2011energy}. The work in \cite{carapellese2014energy} proposes a mixed integer linear programming problem for energy-efficient placement of baseband units in wavelength division multiplexed aggregation networks. Furthermore, optimization and control studies \cite{xiao2021energy} consider a particular model or objective functions specific to the problem formulation without addressing the details of the energy use of networks and their underlying equipment in operational configurations. Understanding energy use in a deployed network context is particularly important to determine how the equipment configurations and architectures impact the overall network energy use and to estimate the power consumption of current operational networks. Recent O-RAN implementations have been limited to just a few open radio units (O-RUs) fanning out from each open distributed unit (O-DU). Furthermore, the compute resources for baseband processing (BBP) can involve multiple cores per 10 Gb/s O-RU connection \cite{Intel2022OpenRAN}. These factors can have a large impact on the energy efficiency of the overall network including transport and compute.

The most commonly used methods for network energy use modeling are the so called `top-down' and `bottom-up' methods. Such analytical models were recently used to analyze and optimize the power consumption of network elements in a fog cloud architecture for 6G networks \cite{yosuf2021cloud}. Yet another approach is the transaction-based modeling technique which models data traffic flow through a network based on mean quantities such as the mean number of hops and typical equipment configurations. This type of modeling has been employed to evaluate the energy consumed by a particular service \cite{baliga2010green} and to observe the effect of variations in the traffic volume of the network \cite{kilper2010power}. Transaction-based models have also been used to calculate optical IP network power \cite{baliga2009energy} and different access networks in wired and wireless systems \cite{baliga2008energy}.

 In this work, we develop transaction-based power consumption models for recent architectures of O-RANs. We formulate the processing power and transmission power individually, then aggregate them to determine the total power consumption when BBP is carried out at various nodes within the network. For 5G, evolved common public radio interface (eCPRI) is used, which includes the digitized radio signals along with control and management data, synchronization signals, and other auxiliary information \cite{pfeiffer2015next}. We also analyze how changes in coverage and network topology impact the power consumption of the O-RAN network through such models. This comprehensive analysis provides critical information that aids in the design and optimization of O-RAN architectures.
\section{O-RAN Architecture and Power Consumption Model}\label{Section II}
In this section, we first describe the O-RAN architecture and then present the proposed transaction-based model for the calculation of power consumption in O-RAN networks. 

\subsection{Network Architecture}
O-RAN specifies standardized open interfaces and constructs within a radio access network (RAN) with the aim of facilitating co-existence among ecosystem participants and creating possibilities for new entrants \cite{oran:online}. The O-RAN architecture provides a highly flexible multi-vendor network by interface standardization and disaggregation of the hardware and software. Fig. \ref{fig:oran_schematics} illustrates an overview of the C-RAN architecture commonly used with O-RAN divided into three levels, namely, access network, metro network and longhaul network. The access network represents the user edge side of the network consisting of O-RUs, and connects the users to the next level of hierarchy through the fronthaul link. In the metro network, the traffic from a cluster of O-RUs is aggregated through the transport equipment in the fronthaul links that connect O-RUs and O-DUs, as depicted in Fig. \ref{fig:oran_schematics}. The open central unit (O-CU) is the next node in the hierarchy of the O-RAN network and it gathers traffic from a set of connected O-DUs and communicates with the core of the network through the high capacity core routers. O-DUs connect to the O-CUs via midhaul links or metro backhaul links if BBP has been fully implemented prior to the O-CU. The longhaul network connects large, often warehouse-scale, DCs to perform computationally intensive tasks and also benefits from the highest efficiency computing resource sharing. The high capacity edge router aggregates the traffic from the network through WDM fiber links and connects the metro network to the core network through backhaul links. WDM links are also used to interconnect core nodes through the gateway router and perform routing for the complete network, as illustrated in Fig. \ref{fig:oran_schematics}.

\subsection{Proposed Power Consumption Model}
We now describe our model to investigate the power consumption in O-RAN architectures using a transaction-based approach \cite{baliga2011energy}. In this approach, we estimate the equipment that would need to be provisioned in order to support a given amount of user traffic or radio coverage. In general, most equipment in use today show little variation in energy use with the traffic levels, typically below 20\% \cite{mellah2011routers}. For this analysis, the equipment is taken to operate at full power, independent of load. In terms of the power per user or user transaction, at baseband, the data is multiplexed with other data and therefore the energy use is in proportion to the transaction data rate and the energy per bit of the equipment in its deployed configuration. In the case of fronthaul eCPRI traffic, the digitized radio waveforms are transported without such baseband multiplexing. Here, we use the common 7.2 phy split, which in principle allows for varying traffic load based on the number of radio blocks in use. However, the equipment needs to be provisioned to handle the maximum load and, therefore, the energy benefits are not realized unless the equipment is oversubscribed or powered off. Some advantage might also be gained from performing processing at other phy splits on intermediate nodes (typical of the 'midhaul' case). These considerations will be the subject of future work.

We focus on the impact of full BBP at different nodes and do not analyze the oversubscription cases although they are allowed for in these models. In practice, such equipment is provisioned to handle a target peak traffic level and, therefore, provisioned above the mean traffic levels by a specific factor. Some consideration is given to allow for traffic growth over time and natural diurnal or other periodic fluctuations. Here, the mean traffic volume is taken over a suitably long time to average such variations, typically on the order of a week or longer \cite{kilper2010power}. The power consumption of deployed equipment also includes various overhead factors such as the power used in cooling and electricity conversion and distribution losses \cite{baliga2011energy}. Another factor is included to account for excess equipment provisioning to provide coverage over a wide area to support mobility and infrequent but necessary user access at diverse locations. We test the proposed transactional-based analytical model by changing the network topology and observed the effect of multiplexing gain with better sharing of resources at processing units (refer Section \ref{Section III}.B).

\begin{figure}
\vspace{2mm}
    \centering
    \includegraphics[width=3in]{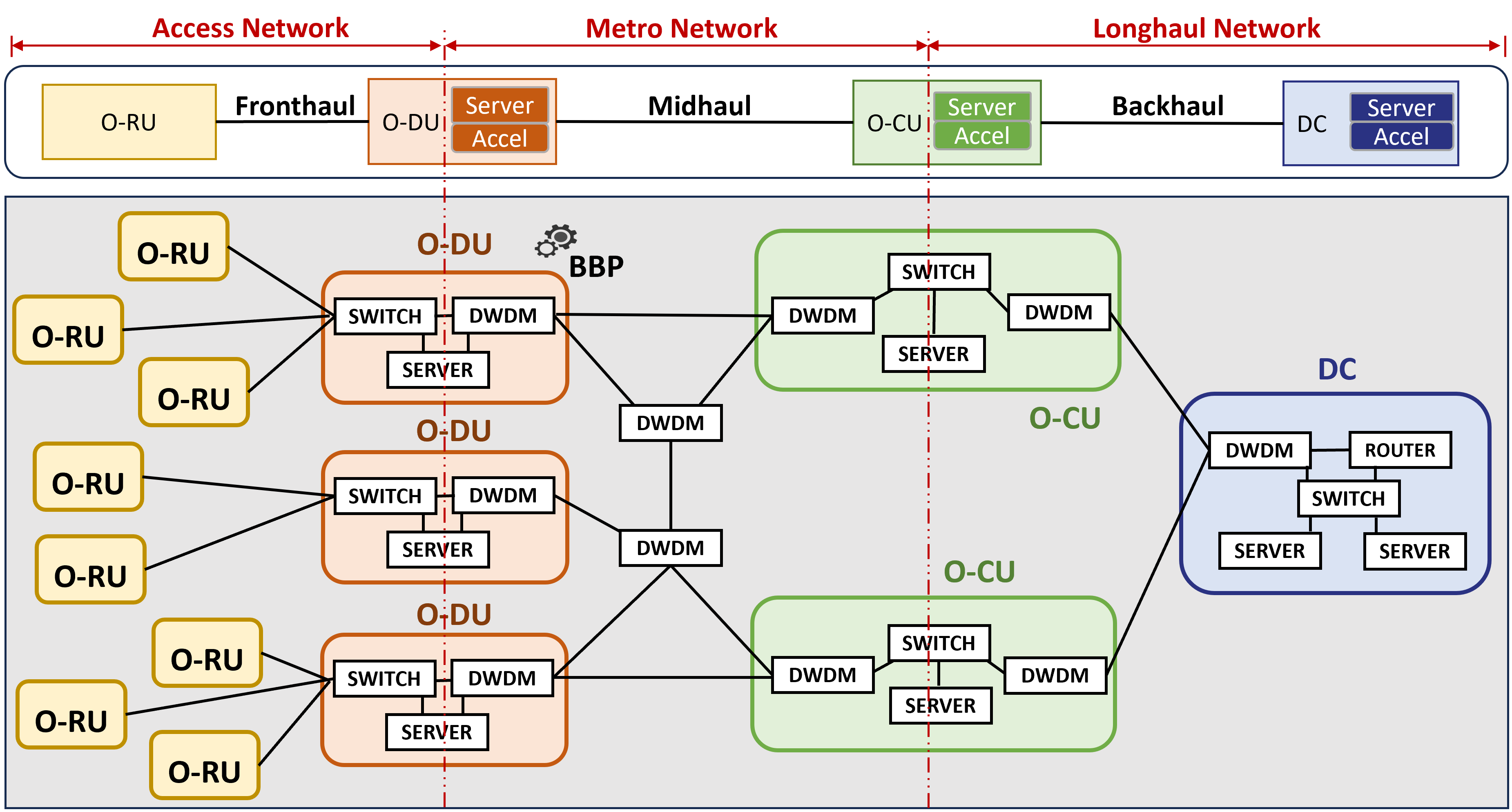}
    \caption{Centralized RAN architecture used in O-RAN showing baseband processing (BBP) at O-DU.} 
    \label{fig:oran_schematics}
\end{figure}

The power consumption model depends on the type and location of processing performed on the eCPRI data. Network equipment consumes power as it processes or transports data traffic traveling from one node to another. This power consumption depends on the amount of data traffic, the type of operations performed on the data, and the respective type of network equipment involved. As discussed earlier, the O-RAN model is comprised of an access network, a metro network, and a core network. These network segments are connected through fronthaul, midhaul, and backhaul transmission links (refer Fig. \ref{fig:oran_schematics}). Besides data transmission, different operations are performed on the user traffic at each node in the network. These operations include BBP and application processing. In this analysis, we only consider BBP from the 7.2 split at the O-RU and explore the impact of performing this BBP at different locations. Note that we include the case of BBP at the O-RU as a limiting case, although this is not done in practice (unless the O-DU and/or O-CU is co-located with the O-RU) and therefore we provide an estimate for such a case if one were to place additional processing with the typical O-RU equipment.  The total power consumption includes the processing power at the network nodes and the transmission power consumed by the network equipment to handle the flow of user traffic over the transmission links. These power components are explained in detail in the following subsections.

\subsubsection{Processing Power} 
The processing power ($P_{pr}$) at any node in an O-RAN network architecture includes two components: the BBP power ($P_{BB}$) and the power ($P_{n}$) consumed by the network equipment at the processing node. Hence, $P_{pr}=P_{BB}+P_{n}$. The first component $P_{BB}$ models the power consumption of the processing server and is present only if BBP is being performed at the node, else $P_{BB}=0$ as the signal passes transparently through the node without processing. The second component $P_{n}$ models the power consumption of the node transport equipment and is governed by the location of the node with respect to the BBP node location. If BBP has not been performed at a previous node then $P_{n}$ depends on the node coverage factor ($\rho_{n}$) and the network eCPRI traffic ($C_{n}$) passing through the node. On the other hand, if BBP has already been performed then $P_{n}$ is governed by the baseband traffic ($C_{u}$) of the user. Once the signal is reduced to baseband, it is transported as multiplexed packet traffic sharing the network equipment with other traffic independent of the radio coverage and, therefore, is independent of the node coverage factor. The processing power is expressed as
\begin{equation} \label{eq:3}
    P_{pr}= 
\begin{cases}
    \alpha_{n}\sigma_{n}\rho_{n}C_{n}\dfrac{ P_{i}}{C_{i}},& \text{before BBP}\\
    \alpha_{n}\sigma_{n}\rho_{n}C_{n}\left(\dfrac{M_{c}P_{c}}{C_{c} }+\dfrac{P_{i}}{C_{i} }\right),              & \text{BBP on node}\\
    \alpha_{n}\sigma_{n}C_{u}\dfrac{ P_{i}}{C_{i}},& \text{after BBP}
\end{cases}
\end{equation}
where $\alpha_{n}$, $\sigma_{n}$, and $\rho_{n}$ are the overprovisioning factor, overhead factor and coverage factor, respectively, of the node. Moreover, $P_i$ and $C_i$ are the power consumption and capacity, respectively, of the network interface card (NIC) or line system chassis present in the node equipment, $M_c$ is the number of server cores and $P_{c}$ and $C_{c}$ are the power consumption and capacity, respectively, of each server core. The number of cores is provisioned based on configurations used for O-RAN radio intelligent controller (RIC) \cite{wu2024hsadr} implementations for a given amount of eCPRI data. The corresponding power per capacity is applied to the fronthaul traffic rate which is multiplied by the coverage factor as described above. The values of the node modeling parameters, $\alpha_{n}$, $\sigma_{n}$, and $\rho_{n}$, depend on the node type as listed in Table \ref{tab1}, where $N_{\text{RU}}$, $N_{\text{DU}}$, $N_{\text{CU}}$, $N_{\text{DC}}$, and $N_{u}$ denote the number of O-RUs, O-DUs, O-CUs, DCs and users, respectively. The number of users is taken as an average over a wide area.

\begin{table}[t!]
\vspace{2mm}
     \caption{\textsc{Modelling Parameters}}
    \begin{center}  
       \begin{tabular}{>{\centering\arraybackslash}p{3.7cm}|>{\centering\arraybackslash}p{0.95cm}|>{\centering\arraybackslash}p{1.4cm}|>{\centering\arraybackslash}p{1cm}}
    \hline
        Network  Segment & Overhead  Factor ($\sigma_{n}$)& Over-provisioning Factor      ($\alpha_{n}$)& Coverage  Factor  ($\rho_{n}$)\\
       
     \hline
         Open Radio Unit (O-RU)& 1 & 5 &$ N_{\text{RU}}/N_{u}$\\
         Open Distribution Unit (O-DU)& 2& 5 &$ N_{\text{DU}}/N_{u}$\\
         Open Central Unit (O-CU)& 2& 5 &$ N_{\text{CU}}/N_{u}$\\
         Data Center (DC)& 1.5& 1.3 &$ N_{\text{DC}}/N_{u}$\\
         Fronthaul Link& 2 & 5&$ N_{\text{RU}}/N_{u}$\\
         Midhaul Link& 2& 5 &$ N_{\text{DU}}/N_{u}$\\
         Backhaul Link& 1.5& 2&$ N_{\text{CU}}/N_{u}$ \\
          \hline 
         \end{tabular}
    \label{tab1}
    \end{center}
\end{table}
\subsubsection{Transmission  Power}
The transmission power in O-RAN includes the power consumed by the UE to send the traffic through the wireless links to an O-RU and from the O-RU to DC through transport equipment like Ethernet switches, routers and optical fiber transmission systems. As the focus of this work is on the transport portion of the network, the power consumed by the UE assumes a simple mean energy per bit, independent of the distance to the antenna and other interference effects.  The power of the UE is therefore expressed as $C_{u}E_{tr}$, where $C_{u}$ is the user data traffic and $E_{tr}$ is the energy per bit of the UE \cite{hribar2021analyse}. We model the power consumption in transmission links and transport equipment like Ethernet switches, routers and WDM transmission systems along the path of the traversing traffic. Hence, the transmission power is expressed as
\begin{equation} \label{eq:8}
\begin{aligned}
    P_{tr}=C_{u}E_{tr} +\alpha_{l}\sigma_{l}\varepsilon_{l}\left[(H_{sh}+1)\frac{P_{s}}{C_{s}}+(H_{lh}+1)\frac{P_{l}}{C_{l}}\right.\\
    \left.+\gamma(H_{rh}+1)\frac{P_{r}}{C_{r}}\right]     
\end{aligned}   
\end{equation}
where $\alpha_{l}$ is the overprovisioning factor of the WDM link and $\sigma_{l}$ is the overhead factor of the link. Here, $H_{sh}$, $H_{lh}$, and $H_{rh}$ are the number of hops across switches, WDM links, and routers, respectively. Similarly, $P_{s}$, $P_{l}$ and $P_{r}$ are the power consumption in the switches, links, and routers, respectively, and $C_{s}$, $C_{l}$ and $C_{r}$ are the corresponding capacities. The factor $\varepsilon_{l}$ accounts for the effect of BBP. Specifically, we use $\varepsilon_{l}=\rho_{l}C_{l}$ before BBP and $\varepsilon_{l}=C_{u}$ after BBP, where $\rho_{l}$ is the link coverage factor and $C_{l}$ is the network eCPRI traffic traversing the link. The values of the link modelling parameters, $\alpha_{l}$, $\sigma_{l}$ and $\rho_{l}$, depend on the link type as listed in Table \ref{tab1} \cite{kilper2010power}. Moreover, for fronthaul and midhaul, we consider only switches and IP/WDM links but for backhaul, we also include routers in the model to cater to the high capacity communication that occurs over backhaul links. This is accounted by the factor $\gamma$ such that $\gamma=1$ for backhaul links, else $\gamma=0$. Using the power consumption model presented in this section, the total power per user consumed by an O-RAN network is calculated as $P_{T}=P_{pr}+P_{tr}$.
\begin{table}[t!]
     \caption{\textsc{Network Equipment}}
    \begin{center}  
    \begin{tabular}{>{\centering\arraybackslash}p{2.5cm}|>{\centering\arraybackslash}p{1.9cm}|>{\centering\arraybackslash}p{1.4cm}|>{\centering\arraybackslash}p{0.9cm}} 
    \hline
        Network\linebreak Devices & Equipment Model & Rated Power (W) & Capacity (Gbps) \\
        \hline
       Router \cite{Cisco80036:online} & Cisco 8000 & 172 &3200 \\
     
        Core Switch \cite{CiscoCat90:online}& Cisco 9600 & 3000  & 25600\\
   Access Switch \cite{CiscoCat90:online}& Cisco 1300 & 86.7  & 480\\
         Link \cite{1FINITY™41:online} & 1FINITY T600& 4265 & 9600\\
          Radio \cite{BenetelR56:online} & Benetel 650& 110 & 22\\
           \hline
    \end{tabular}
    \label{tab2}
    \end{center}
    \end{table}

\begin{figure*}[!t]
    \centering
     \vspace{2mm}
    \includegraphics[width=6in]{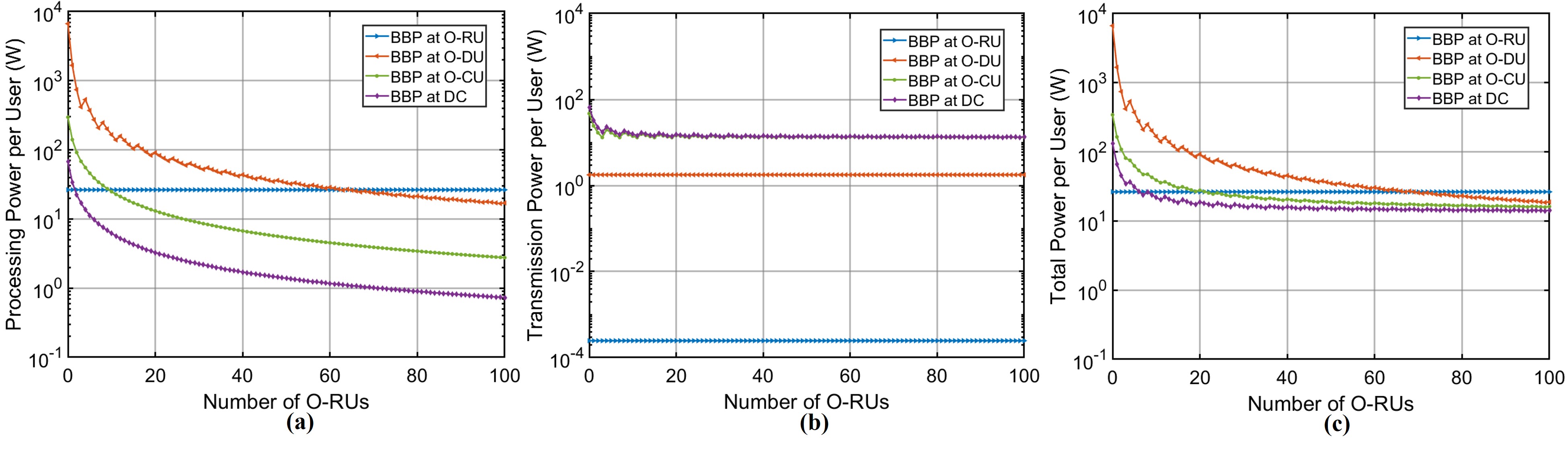}
    \vspace{-3mm}
    \caption{Variation in (a) processing, (b) transmission, and (c) total power consumption per user when the number of O-RUs is increased with baseband processing (BBP) at different nodes mentioned in the legend. Here, the number of users per O-RU is 10.} 
    \label{fig:power_vs_cov}
\end{figure*}
 
\section{Results}\label{Section III}
In this section, we present the results obtained using the proposed model described in Section \ref{Section II}.B to analyze the power consumption of O-RANs considering different centralized architectures. Details of the network equipment used in our analyses are given in Table \ref{tab2}. We assume that the average monthly data consumption of a user is 10 Gigabytes. Typical of O-RAN deployments, the servers at O-RU, O-DU, and O-CU have 4 cores with a total BBP capacity of 1 Gbps, whereas the server at DC has 20 cores which gives a cumulative BBP capacity of 5 Gbps per server. \cite{Acceller65:online}. The thermal design power per core for the servers at O-RU, O-DU and O-CU is 6 W \cite{IntelXeo93:online}, and for those at DC, it is 5.5 W \cite{IntelXeo9:online}. The energy expended in the UE is $E_{tr}=$ 25 nJ/bit \cite{hribar2021analyse}. These design parameters are fixed unless mentioned otherwise. We obtain the processing power and transmission power separately and combine these to obtain the total power consumption for different BBP locations. Moreover, we study the effect of change in the number of O-RUs and network topology on the power consumption in O-RANs.

\subsection{Effect of Number of O-RUs}\label{Subsection A}
We vary the number of O-RUs and evaluate the power consumption of the network when BBP is performed at different O-RAN nodes, namely O-RU, O-DU, O-CU and DC. For the outdoor 5G stand-alone O-RU considered here (refer Table \ref{tab2}), the corresponding fronthaul traffic based on the 7.2 split is 11 Gbps \cite{virtualization2016functional}. As described above, we assume that the radio network is provisioned to support the maximum eCPRI rate supported for each O-RU. Here, the number of users per O-RU is fixed at 10. Moreover, in order to model the impact of fanout limitations of O-RAN, each O-DU is provisioned to serve a maximum of four O-RUs. When the number of O-RUs exceeds a multiple of 4, another O-DU is added to the system.

In Fig. \ref{fig:power_vs_cov}, we depict the variation in transmission, processing and total power consumption per user as the number of O-RUs is increased from 1 to 100 with BBP being performed at different O-RAN nodes. For this constant number of users per O-RU case, perhaps counter-intuitively the power per user decreases as the number of O-RUs increases. Clearly, if the number of users were constant and the number of O-RUs were increased this would not be the case. We observe in Fig. \ref{fig:power_vs_cov}(a) that when the number of O-RUs is less, the processing power for BBP at O-DU, O-CU and DC are higher than that at O-RU because the resources at O-DU, O-CU and DC are not efficiently utilised. However, when the number of O-RUs is incremented, the resources at O-DU, O-CU and DC are efficiently used, so processing powers for BBP at the O-DU, O-CU and DC decrease. We also observe that the power consumption for BBP at O-CU and DC decreases smoothly because the number of O-RUs in the network is being increased linearly. However, the corresponding variations for BBP at O-DU are jagged due to the inclusion of an additional O-DU for every four O-RUs. When the O-DUs are fully utilized, the power consumption per user is lower. Additionally, for the same number of O-RUs, power consumption in performing BBP at DC is lower than that at O-CUs because the servers at DC have higher capacity. 

We observe a reverse trend in the case of transmission power, as illustrated in Fig. \ref{fig:power_vs_cov}(b), because the transmission power greatly depends upon the number of hops over which the fronthaul traffic is transported since it is many times larger than the baseband traffic and does not benefit from multiplexing with other traffic. When we perform BBP at O-RU (which is the edge node), the amount of traffic traversing further into the network is considerably reduced. Hence, the transmission power consumed in this case is much lower than that consumed when BBP is performed at the DC (which is located in the network core) because in the latter case, a large amount of eCPRI data has to travel to the network core for BBP. As the network equipment must be provisioned to account for traffic variations and does not benefit from sleep modes or savings due to low traffic periods, as typical for most equipment today, the benefits of the 7.2 splits investigated previously are not realized. This results in consuming high transmission power. The processing power and transmission power are combined to obtain the total power consumption per user as depicted in Fig. \ref{fig:power_vs_cov}(c). For multi-core processing requirements in emerging O-RAN deployments for such transport network scenarios, performing BBP at the DC is the most power-efficient due to its benefit from the highly efficient and shared servers, as processing power dominates the transmission power consumed by the user traffic. If we increase the fanout of O-RU and O-DU nodes to support greater densification and allow for greater centralization and sharing of the processing resources, the relative power per user for processing may decrease. Therefore, we now investigate the impact of the nodal fanout.

\begin{figure*}[!t]
    \centering
    \includegraphics[width=6in]{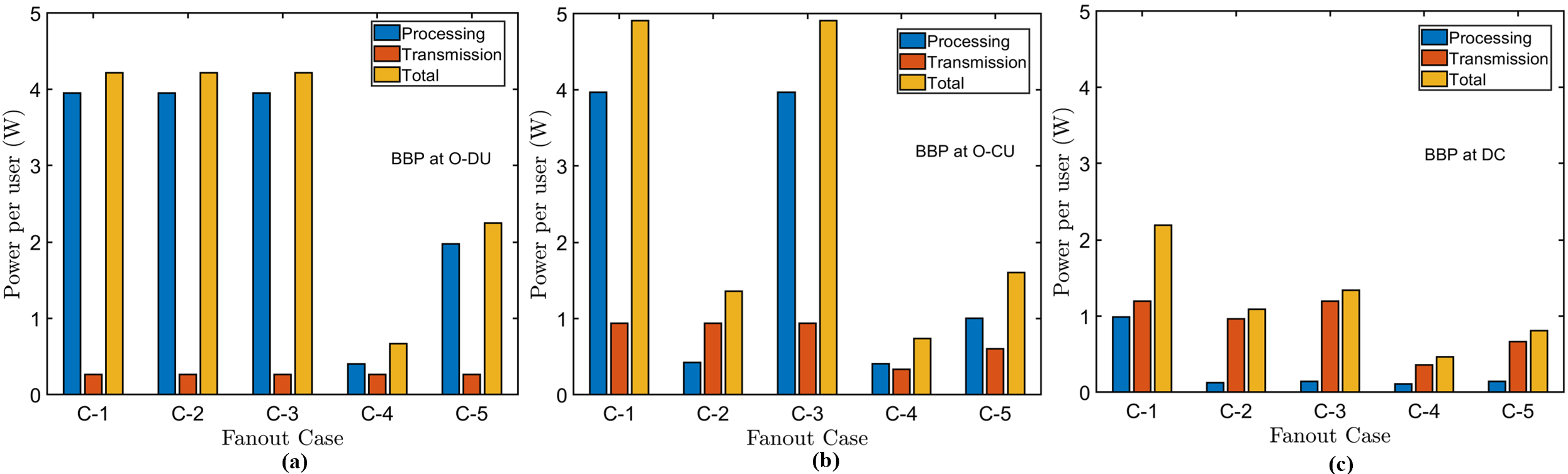}
   \vspace{-3mm}
    \caption{Variation in power consumption per user for five different cases of nodal fanout (refer Table \ref{tab3}) with baseband processing (BBP) at (a) O-DU, (b) O-CU, and (c) DC. Here, the number of users per O-RU is 10.} 
    \label{fig:diff_fan_tot_pow}
\end{figure*}

\begin{figure*}[!t]
    \centering
    \includegraphics[width=6in]{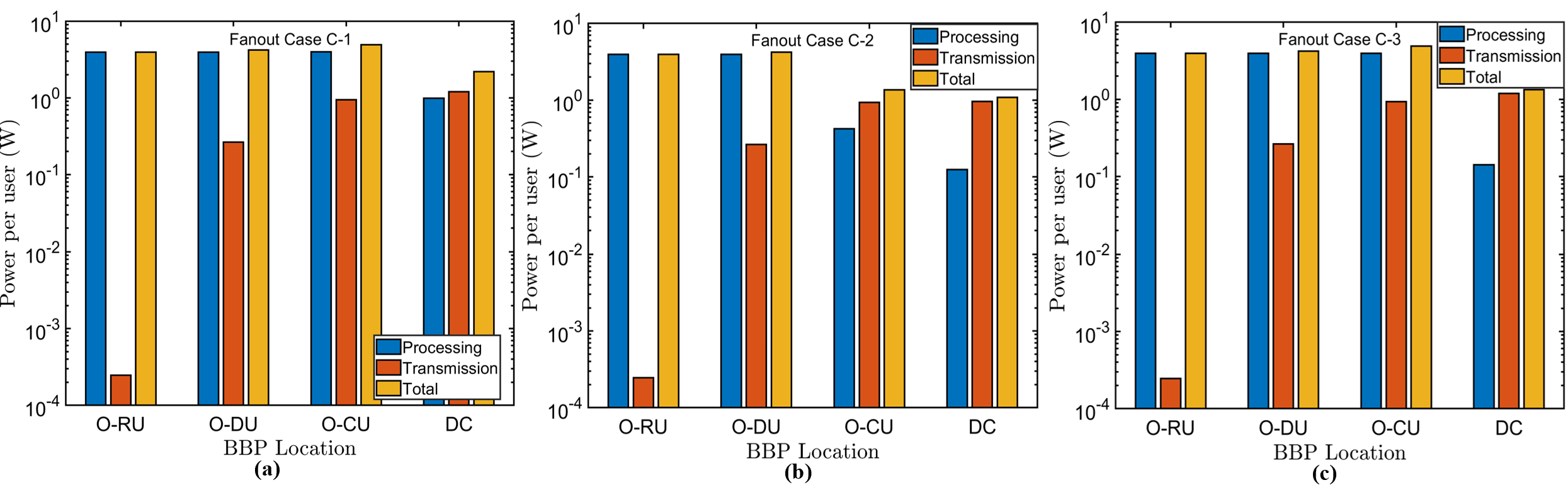}
    \vspace{-3mm}
    \caption{Variation in total power consumption per user with baseband processing (BBP) at different nodes when the nodal fanout case (a) C-1, (b) C-2, and (c) C-3. Here, the number of users per O-RU is 10.}
    \label{fig:diff_bbp_tot_pow}
\end{figure*}

\subsection{Effect of Nodal Fanout}
Nodal fanout quantifies the number of `child' nodes connected to a `parent' node, and can be calculated as the ratio of number of successive nodes. For example, the O-DU fanout is obtained by dividing the number of O-RUs by the number of O-DUs, and so on. We test five different cases of nodal fanout options, as listed in Table \ref{tab3}, to emulate a diverse range of practical configurations in an O-RAN architecture. For each case, we obtain the power consumed when BBP is performed at different nodes in the network with the number of users per O-RU fixed at 10. The power consumption when BBP is performed at the O-RU location considers the transmitted data to be fully at baseband and therefore benefits from conventional packet multiplexing. Hence, the power consumption when BBP is performed at RU depends on the number of users but has no contribution from any of the other O-RAN nodes. So, the power consumed (when BBP is at O-RU) is the same for each fanout case and is not discussed here.

The power consumption for BBP at the O-DU is shown in Fig. \ref{fig:diff_fan_tot_pow}(a). Here, the transmission power is constant but for higher O-DU fanout (C-4 and C-5), the processing power is lower due to more efficient utilization of the O-DUs. Consequently, BBP at the O-DU is more power-efficient for higher O-DU fanouts. In Fig. \ref{fig:diff_fan_tot_pow}(b), we depict the power consumption when BBP is done at the O-CU. Here, the results are affected by the fanouts of both the O-DU and O-CU. In C-2, the O-CU fanout is high, thereby reducing the processing power. In C-4, the O-DU fanout is high and so the transmission power is lower than in the other cases because of the coverage gain at the O-DU. Fig. \ref{fig:diff_fan_tot_pow}(c) shows the power consumption for the fanout cases when BBP is performed at the DC. The fanouts of O-DU, O-CU and DC affect the power consumption in this case. Here, the transmission power is always higher than the processing power per user. The transmission power is lower in cases with high O-CU fanout (case C-2) and high O-DU fanout (cases C-4). In Fig. \ref{fig:diff_bbp_tot_pow}, we examine the power per user for C-1, C-2 and C-3, individually. For C-1, the number of units at each node is the same. So, BBP at O-RU, O-DU and O-CU consume the same processing power (Fig. \ref{fig:diff_bbp_tot_pow}(a)). However, for BBP at the DC, the processing energy is low due to the efficient and shared computing resources at the DC. Additionally, the transmission power increases if we shift BBP farther away from the edge as the eCPRI data has to travel a larger distance. In C-2, the O-CU fanout is higher so the processing power for BBP at the O-CU is much lower, as depicted in Fig. \ref{fig:diff_bbp_tot_pow}(b). For C-3, the DC fanout is largest and hence, the processing power consumed for BBP at DC is very low as shown in Fig. \ref{fig:diff_bbp_tot_pow}(c).
\begin{table}[!t]
     \caption{\textsc{Different Cases of Nodal Fanout}}
    \begin{center}  
       \begin{tabular}{>{\centering\arraybackslash}p{1cm}|>{\centering\arraybackslash}p{1.6cm}|>{\centering\arraybackslash}p{1.6cm}|>{\centering\arraybackslash}p{1.5cm}}
    \hline
        Fanout Case&  O-DU Fanout ($N_{\text{RU}}/N_{\text{DU}}$)& O-CU Fanout ($N_{\text{DU}}/N_{\text{CU}}$)& DC Fanout ($N_{\text{CU}}/N_{\text{DC}}$)\\
     \hline
       C-1 & 1& 1& 1 \\ 
       C-2 & 1& 10& 1 \\  
       C-3  &1 & 1& 10 \\ 
       C-4 & 10& 1& 1  \\  
       C-5  & 2& 2& 2 \\  
          \hline 
         \end{tabular}
    \label{tab3}
    \end{center}
\end{table}
\section{Conclusion}
We investigate analytical transaction-based models for power consumption in O-RAN architectures using representative equipment and deployment scenarios available today. We evaluate the processing power and transmission power consumed in the network when baseband processing (BBP) is performed at different network nodes. Due to the number of cores used for processing O-RAN signals, the processing power is a strong determinant of the overall network efficiency with efficient data center processing being preferred. This highlights the need to keep the BBP compute requirement for O-RAN low and to make greater use of centralization. As the O-RAN fanout capabilities are improved, however, the high fronthaul transmission capacity requirements will become a factor once again. Performing BBP at intermediate O-DU/O-CU nodes ensures that transmission energy costs are kept low while benefiting from more efficient BBP at high utilization. Looking at the two extremes, when the number of users per O-RU is 10, the total power consumed by each user is reduced by $\sim$80\%, if the BBP is performed at a DC instead of at an O-RU. The study presented in this paper also shows that the processing power is sensitive to how it is shared across a network and requires consideration for where the processing is done relative to how the node fanout among O-RU, O-DU, and O-CU elements. As the technology matures, we anticipate that these figures will evolve significantly, which can be readily studied and understood through the proposed transaction-based energy model approach. We further plan to explore the latency limits and the effect of variable computational resources and network equipment in O-RAN. 

\section*{Acknowledgments}
\small{This work was supported in part by Reasearch Ireland grants 18/CRT/6222 and 13/RC/2077\_P2; and EU MSCA grant 101155602.}

\bibliographystyle{ieeetr}
\bibliography{references}

\begin{thebibliography}{10}

\bibitem{oran:online}
``{ORAN Alliance, 2023}.'' Accessed: Oct. 16, 2024. [Online]. Available: \url{https://www.o-ran.org/}.

\bibitem{ORANVert0:online}
``{O-RAN Vertical Industries White Paper December 2023}.'' \url{https://mediastorage.o-ran.org/white-papers/O-RAN.WG1.Vertical-Industry-White-Paper-2023-12.pdf}.
\newblock (Accessed on 11/01/2024).

\bibitem{baliga2011energy}
J.~Baliga, R.~Ayre, K.~Hinton, and R.~S. Tucker, ``Energy consumption in wired and wireless access networks,'' {\em IEEE Communications Magazine}, vol.~49, no.~6, pp.~70--77, 2011.

\bibitem{carapellese2014energy}
N.~Carapellese {\em et~al.}, ``Energy-efficient baseband unit placement in a fixed/mobile converged wdm aggregation network,'' {\em IEEE J. Sel. Areas Commun.}, vol.~32, no.~8, pp.~1542--1551, 2014.

\bibitem{xiao2021energy}
Y.~Xiao, J.~Zhang, and Y.~Ji, ``{Energy-efficient DU-CU deployment and lightpath provisioning for service-oriented 5G metro access/aggregation networks},'' {\em J. Lightw. Technol.}, vol.~39, no.~17, pp.~5347--5361, 2021.

\bibitem{Intel2022OpenRAN}
{Intel Network Builders}, ``Determining the right hardware for open {RAN} deployment,'' tech. rep., Intel Corporation, 2022.
\newblock [Online]. Available: \url{https://networkbuilders.intel.com/docs}.

\bibitem{yosuf2021cloud}
B.~A. Yosuf {\em et~al.}, ``Cloud fog architectures in {6G} networks,'' in {\em 6G Mobile Wireless Networks}, pp.~285--326, Springer, 2021.

\bibitem{baliga2010green}
J.~Baliga, R.~W. Ayre, K.~Hinton, and R.~S. Tucker, ``Green cloud computing: Balancing energy in processing, storage, and transport,'' {\em Proceedings of the IEEE}, vol.~99, no.~1, pp.~149--167, 2010.

\bibitem{kilper2010power}
D.~Kilper {\em et~al.}, ``Power trends in communication networks,'' {\em IEEE J. Sel. Topics Quantum Electron.}, vol.~17, no.~2, pp.~275--284, 2010.

\bibitem{baliga2009energy}
J.~Baliga {\em et~al.}, ``Energy consumption in optical {IP} networks,'' {\em Journal of Lightwave Technology}, vol.~27, no.~13, pp.~2391--2403, 2009.

\bibitem{baliga2008energy}
J.~Baliga {\em et~al.}, ``Energy consumption in access networks,'' in {\em Proc. Conference on Optical Fiber Communication}, pp.~1--3, 2008.

\bibitem{pfeiffer2015next}
T.~Pfeiffer, ``Next generation mobile fronthaul and midhaul architectures,'' {\em J. Opt. Commun. Netw.}, vol.~7, no.~11, pp.~B38--B45, 2015.

\bibitem{mellah2011routers}
H.~Mellah and B.~Sanso, ``Routers vs switches, how much more power do they really consume? {A} datasheet analysis,'' in {\em Proc. IEEE International Symposium WoWMoM}, pp.~1--6, 2011.

\bibitem{wu2024hsadr}
F.~Wu {\em et~al.}, ``{HSADR: A} new highly secure aggregation and dropout-resilient federated learning scheme for radio access networks with edge computing systems,'' {\em IEEE Transactions on Green Communications and Networking}, vol.~8, no.~3, pp.~1141--1155, 2024.

\bibitem{hribar2021analyse}
J.~Hribar {\em et~al.}, ``Analyse or transmit: Utilising correlation at the edge with deep reinforcement learning,'' in {\em Proc. IEEE Global Communications Conference}, pp.~1--6, 2021.

\bibitem{Cisco80036:online}
``{Cisco 8000 Series Routers - Cisco}.'' Accessed: Aug. 16, 2024. [Online]. Available: \url{https://www.cisco.com/site/us/en/products/networking/routers/8000-series/index.html}.

\bibitem{CiscoCat90:online}
``{Cisco Catalyst Series Switches - Cisco}.'' Accessed: Aug. 14, 2024. [Online]. Available: \url{https://www.cisco.com/c/en/us/support/switches/category.html}.

\bibitem{1FINITY™41:online}
``{1FINITY™ T600 - Fujitsu Network Communications : Fujitsu United States}.'' Accessed: Aug. 16, 2024. [Online]. Available: \url{https://www.fujitsu.com/us/products/network/products/1finity-t600/}.

\bibitem{BenetelR56:online}
``{Benetel-RAN650-Product-Specification-Sheet-v.1.1.pdf}.'' Accessed: Aug. 03, 2024. [Online]. Available: \url{https://benetel.com/wp-content/uploads/2021/11/Benetel-RAN650- Product-Specification-Sheet-v.1-1.pdf}.

\bibitem{Acceller65:online}
``{Accelleran’s 5G dRAX™ cloud-native Open RAN software}.'' Accessed: Aug. 04, 2024. [Online]. Available: \url{https://accelleran.com/accellerans-5g-drax-cloud-native-open-ran-software-now-available/}.

\bibitem{IntelXeo93:online}
``{Intel Xeon Silver 4216 Processor 22M Cache 2.10 GHz Product Specifications}.'' Accessed: Aug. 01, 2024. [Online]. Available: \url{https://ark.intel.com/content/www/us/en/ark/products/193394/intel-xeon-silver-4216-processor-22m-cache-2-10-ghz.html}.

\bibitem{IntelXeo9:online}
``{Intel Xeon Gold 6262V Processor 33M Cache 1.90 GHz Product Specifications}.'' Accessed: Aug. 01, 2024. [Online]. Available: \url{https://ark.intel.com/content/www/us/en/ark/products/193972/intel-xeon-gold-6262v-processor-33m-cache-1-90-ghz.html}.

\bibitem{virtualization2016functional}
``{Small Cell Virtualization: Functional splits and use cases},'' in {\em Small Cell Forum Release}, vol.~6, p.~55, 2016.

\end{thebibliography}

\end{document}